\documentclass[prb,reprint,floatfix]{revtex4-1}
\usepackage{siunitx}
\usepackage{graphicx}
\usepackage{xcolor}
\usepackage{amsmath}

\newcommand{\mrx}{Mn$_2$Ru$_x$Ga}
\newcommand{\mrg}{MRG}
\newcommand{\tcmp}{T_{\text{comp}}}

\begin{document}
\author{G. Bonfiglio}
\affiliation{Radboud University, Institute for Molecules and Materials, 6525 AJ Nijmegen, The Netherlands}
\author{K. Rode}
\author{G.Y.P.~Atcheson}
\author{P. Stamenov}
\author{J.M.D.~Coey}
\affiliation{CRANN, AMBER and School of Physics, Trinity College Dublin, Ireland}
\author{A.V. Kimel}
\author{Th.~Rasing}
\author{A. Kirilyuk}
\affiliation{Radboud University, Institute for Molecules and Materials, 6525 AJ Nijmegen, The Netherlands}
\affiliation{FELIX Laboratory, Radboud University, Toernooiveld 7, 6525 ED Nijmegen, The Netherlands}

\title{Sub-picosecond exchange-relaxation in the compensated ferrimagnet \mrx}

\begin{abstract}
  We study the demagnetization dynamics of the fully compensated half-metallic ferrimagnet \mrx. While the two antiferromagnetically coupled sublattices are both composed of manganese, they exhibit different temperature dependencies due to their differing local environments. The sublattice magnetization dynamics triggered by femtosecond laser pulses are studied to reveal the roles played by the spin and intersublattice exchange. We find a two-step demagnetization process, similar to the well-established case of \mbox{Gd(FeCo)$_3$}, where the two Mn-sublattices have different demagnetization rates. The behaviour is analysed using a four-temperature model, assigning different temperatures to the two manganese spin baths. Even in this strongly exchange-coupled system, the two spin reservoirs have considerably different behaviour. The half-metallic nature and strong exchange coupling of \mrx\ lead to spin angular momentum conservation at much shorter time scales than found for \mbox{Gd(FeCo)$_3$} which suggests that low-power, sub-picosecond switching of the net moment of \mrx\ is possible. 

\end{abstract}

\maketitle


\citet{Beaurepaire1996} demonstrated in 1996 that the magnetization of a ferromagnet can be changed on the sub-picosecond timescale thereby raising the possibility that the combination of magnetism and light may bridge the `terahertz gap' in spin electronic devices.
Further striking observations were made a few years later when short laser pulses were shown to switch the net magnetization of a ferrimagnet. This is now called all-optical switching (AOS)\cite{Stanciu2007,Lambert2014,Stupakiewicz2017}.
The best-studied AOS material is the amorphous compensated ferrimagnet \mbox{Gd(FeCo)$_3$} where toggle switching can be understood by allowing for exchange of angular momentum between the Gd and FeCo sublattices due to exchange interaction between them\cite{Radu2011,Ostler2012,Mentink2012}.
The fundamentals of AOS have been subject to intense investigation since then, and several different models\cite{Mentink2012,Atxitia2010,Cornelissen2016,Gridnev2018} have been put forward, all based on transfer of energy and angular momentum between the electronic, lattice and spin subsystems. In order to understand the dynamics of switching, exchange, electron-phonon interaction and spin-lattice relaxation must all be considered\cite{Gridnev2018}. The key feature of all the models however, is different relaxation dynamics for the two sublattices. The Gd and transition metal spin reservoirs must be described separately, adding one extra temperature\cite{Mekonnen2013} to the widely accepted, phenomenological, three-temperature model (3TM)\cite{Beaurepaire1996} for a ferromagnet. This four-temperature description (4TM), reproduces the demagnetization dynamics of these alloys quite well, indirectly validating different demagnetization dynamics for the two sublattices, as was observed by XMCD\cite{Radu2011}.

It was shown by \citet{Mangin2014} that all-optical influence on the magnetization can be achieved in various structures: alloys, multilayers, heterostructure and rare-earth-free synthetic ferrimagnets. From those results it was possible to infer three empirical design rules for a ferrimagnet to show AOS: antiferromagnetic coupling, non-equivalent sublattices and perpendicular anisotropy\cite{Kimel2014}.

In this respect, a yet unexplored and fascinating new material for AOS is the ferrimagnetic half-metal \mrx\cite{Kurt2014} (\mrg). Due to its half-metallicity, it could be an ideal material for spintronic devices\cite{Borisov2016,Borisov2017,Thiyagarajah2015}. The two antiferromagnetically coupled sublattices, both Mn-based, present different temperature dependencies of their magnetizations due to the different local electronic environment at the two different crystallographic sites ($4c$ and $4a$ positions in the $F\bar{4}3m$ space group)\cite{Betto2015}. There is a spin gap in one sublattice of about \SI{1}{\electronvolt}, and the electrons at the Fermi level in the other subband belong predominantly to the $4c$ sublattice\cite{Kurt2014}. We have shown that the laser-induced spin precession resembles that of a ferromagnet, but with much higher frequency and relatively low damping\cite{Bonfiglio2019}.

Recently, \citet{Banerjee2019} have shown that \mrg\ exhibits single-pulse all-optical toggle switching that is both similar to and very different from \mbox{Gd(FeCo)$_3$}.
In particular, the two Mn sublattices are strongly (compared to \mbox{Gd(FeCo)$_3$}) exchange-coupled\cite{Fowley2018} and of the same magnitude. Unlike \mbox{Gd(FeCo)$_3$}, the differing sublattice demagnetisation rates cannot be determined entirely by the sublattice moments and their angular momenta. It must be driven by angular momentum \emph{conservation} and intersublattice exchange relaxation. The question is if the spin-resolved heat capacity, determined almost entirely by the spin-polarized density of states at the Fermi level, is sufficiently distinct to account for the substantial difference in characteristic demagnetisation times. Are the two spin reservoirs in equilibrium with each other during the entire de- and re-magnetization?

In order to answer this question, we study the demagnetization dynamics of \mrg\, in applied magnetic fields of up to \SI{7}{\tesla}. The initial ultrafast (less than one picosecond) demagnetization is followed by a plateau or a remagnetization, and a slower demagnetization process after this. We will show that numerical simulations based on the 4TM reproduce the experimental data, and provide us with a set of intrinsic material parameters that help understand the ultrafast behaviour of \mrg. The relatively strong inter-sublattice exchange interaction leads to overall faster dynamics of \mrg\ than has been observed for Gd(FeCo)$_3$.


The ferrimagnetic \mrx\ sample used in these experiments has a magnetic compensation point $\tcmp \sim \SI{250}{\kelvin}$ and the Curie temperature is $T_{\text{C}} \sim \SI{550}{\kelvin}$. Thin films of \mrg\ were grown on MgO (001) substrates in a `Shamrock' sputter deposition cluster with a base pressure of \SI{2e-8}{Torr}. The substrate was kept at \SI{250}{\degreeCelsius} during deposition of \mrg, and a protective, $\sim \SI{3}{\nano\meter}$, layer of aluminium oxide was added post-deposition at room temperature. Further information on sample deposition can be found elsewhere\cite{Betto2015}. The thickness of the sample is \SI{50}{\nano\metre} and $x = 0.7$.

The demagnetization dynamics were investigated using a two-colour pump-probe scheme in a Faraday geometry inside a $\mu_0 H_{\text{max}} = \SI{7}{\tesla}$ superconducting magnet. Data shown were recorded below $\tcmp$ at \SIlist{210;230}{\kelvin}. Both pump and probe were produced by a Ti:sapphire femtosecond pulsed laser amplifier with a central wavelength of \SI{800}{\nano\metre}, a pulse width of \SI{40}{\femto\second} and a repetition rate of \SI{1}{\kilo\hertz}. The beam was split in two parts, with the high-intensity one frequency doubled by a BBO crystal (producing $\lambda=\SI{400}{\nano\metre}$) and used as a pump pulse. The lower-intensity part with the wavelength of \SI{800}{\nano\metre} acted as the probe. The time delay between the two pulses was adjusted using a mechanical delay stage. To improve the signal-to-noise ratio, the pump pulses were modulated by a synchronized mechanical chopper at \SI{500}{\hertz} for subsequent lock-in detection. Both beams were linearly polarized, and with spot sizes on the sample of \SIlist{150;70}{\micro\metre} for pump and probe, respectively. After interaction with the sample, the probe beam was split in two orthogonally polarized components using a Wollaston prism. The pump-induced changes in transmission and Faraday rotation were thus detected by measuring the sum and the difference in intensity of the two signals. Further, the external magnetic field was applied along the easy axis (i.e.\ out of plane), to fully saturate the sample between each pump pulse.

\begin{figure}
	\centering
	\includegraphics[width=\columnwidth]{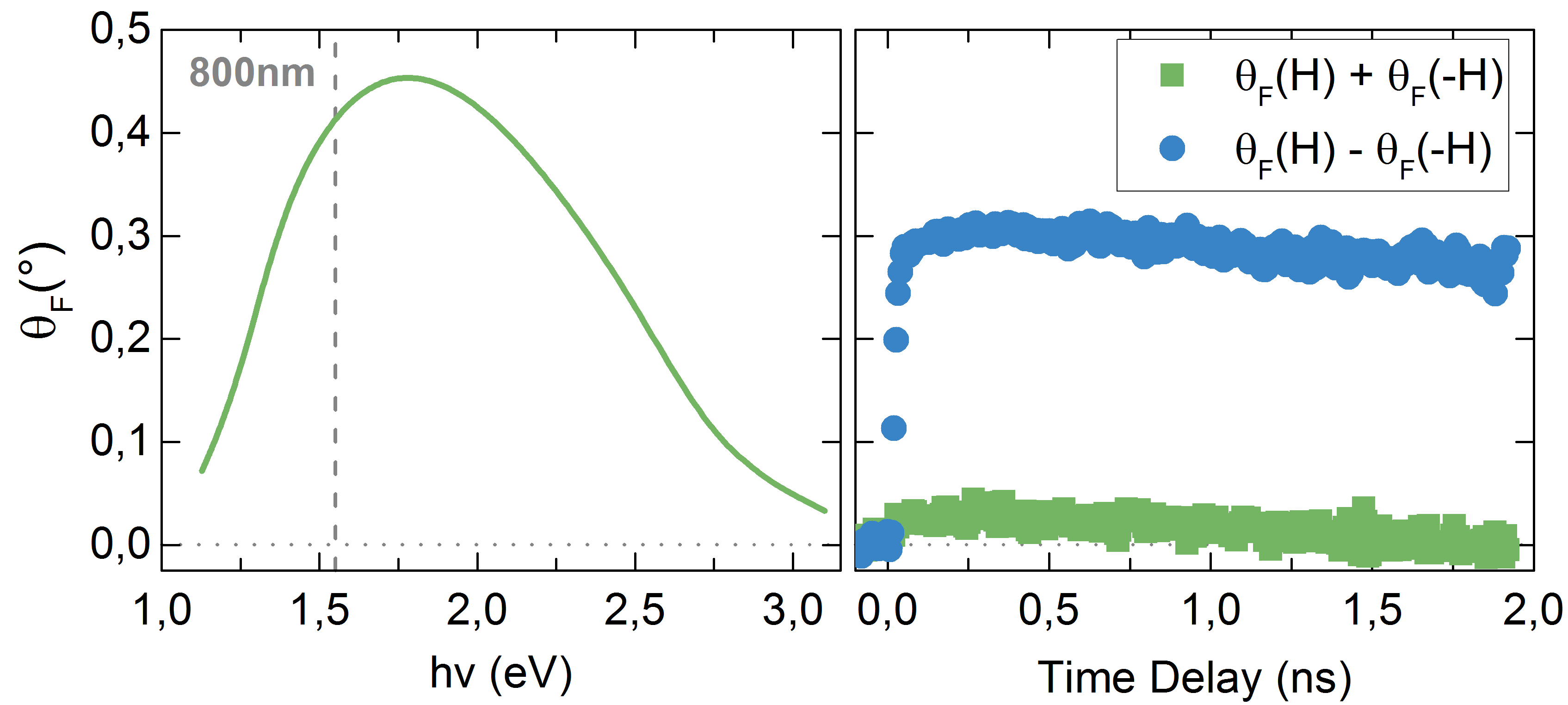}
	\caption{(Left) \mrg\ Faraday rotation ($\theta_F$) as a function of the photon energy. The dashed vertical line indicates the probe energy ($\lambda = \SI{800}{\nano\metre}$) for magnetization dynamics hereafter. (Right) $\theta_F$ dynamics. Green squares show effects that are even in applied field, while blue squares show odd ones. They are the sum and the difference of data obtained with positive and negative applied fields ($\pm\SI{3}{\tesla}$) at $T=\SI{220}{\kelvin}$. Changes in transmission are small and negligible compared to the changes induced by magnetisation dynamics.}
	\label{fig:mo-spectrum_two-field}
\end{figure}

In \figurename~\ref{fig:mo-spectrum_two-field} the magneto-optical spectrum of \mrg\ is plotted from \SIrange{1.12}{3.1}{\electronvolt}. The photons with energy \SI{1.55}{\electronvolt} probe mainly the $4c$ sublattice. The main contribution to the dielectric permittivity in the visible and near infrared arises from the Drude tail\cite{Fleischer2018} so that the magneto-optical probe follows the behaviour of the highly spin-polarised conduction band. However, a rather large density of states in the vicinity of the spin gap indicates that excitation of $4a$ states is possible as well.\cite{Zic2016} We note that the Faraday rotation does not change sign from \SIrange{1}{3}{\electronvolt} (\figurename~\ref{fig:mo-spectrum_two-field}), the hysteresis loops obtained by MOKE/Faraday match those obtained by anomalous Hall effect\cite{Bonfiglio2019}, and the anomalous Hall angle ($\rho_{xy}/\rho_{xx}$) almost perfectly matches the same ratio extrapolated from the optical measurements\cite{Fleischer2018}. This indicates that the sublattice contributing most of the electrons close to the Fermi level is mostly responsible for the magneto-optical response at \SIlist{800; 400}{\nano\metre}.

The effect of a pump pulse with the fluence of \SI{6.5}{\milli\joule\per\centi\metre\squared} is shown in \figurename~\ref{fig:mo-spectrum_two-field}. Here, the demagnetization process for long time scales, up to \SI{2}{\nano\second}, exhibits opposite signs for opposite fields. In the dynamics we can distinguish effects that are odd and even in magnetic field from the difference and the sum (\figurename~\ref{fig:mo-spectrum_two-field}). The difference is assigned to the magnetization dynamics while the sum can be explained by time-dependent changes of transmission through the sample.

AOS generally proceeds in three different steps. First the ultrashort laser pulse leads to a drastic increase of the electronic temperature, above the magnetic ordering temperature $T_{\text{C}}$. Subsequently, heat is transferred from the hot electrons to the spin subsystem in around \SI{1}{\pico\second}, leading to rapid demagnetization. In the case where the atomic moments of the two sublattices are substantially different, as for \mbox{GdCo}, they will demagnetize with different characteristic times, proportional to $\mu_i/\alpha_i$, where $\mu_i$ is the sublattice atomic moment and $\alpha_i$ its damping constant. A transient ferromagnetic state arises, followed by complete switching of the magnetic order. An important part of this process is that angular momentum is exchanged between the sublattices, due to exchange relaxation\cite{Mentink2012}, resulting in acceleration of the demagnetization for both sublattices.

In our experiment, the strong field  applied along the easy axis ensures that when the system cools down (starting from few hundreds of picoseconds, \figurename~\ref{fig:mo-spectrum_two-field}) the initial magnetic state is restored.

\begin{figure}
	\centering
	\includegraphics[width=\columnwidth]{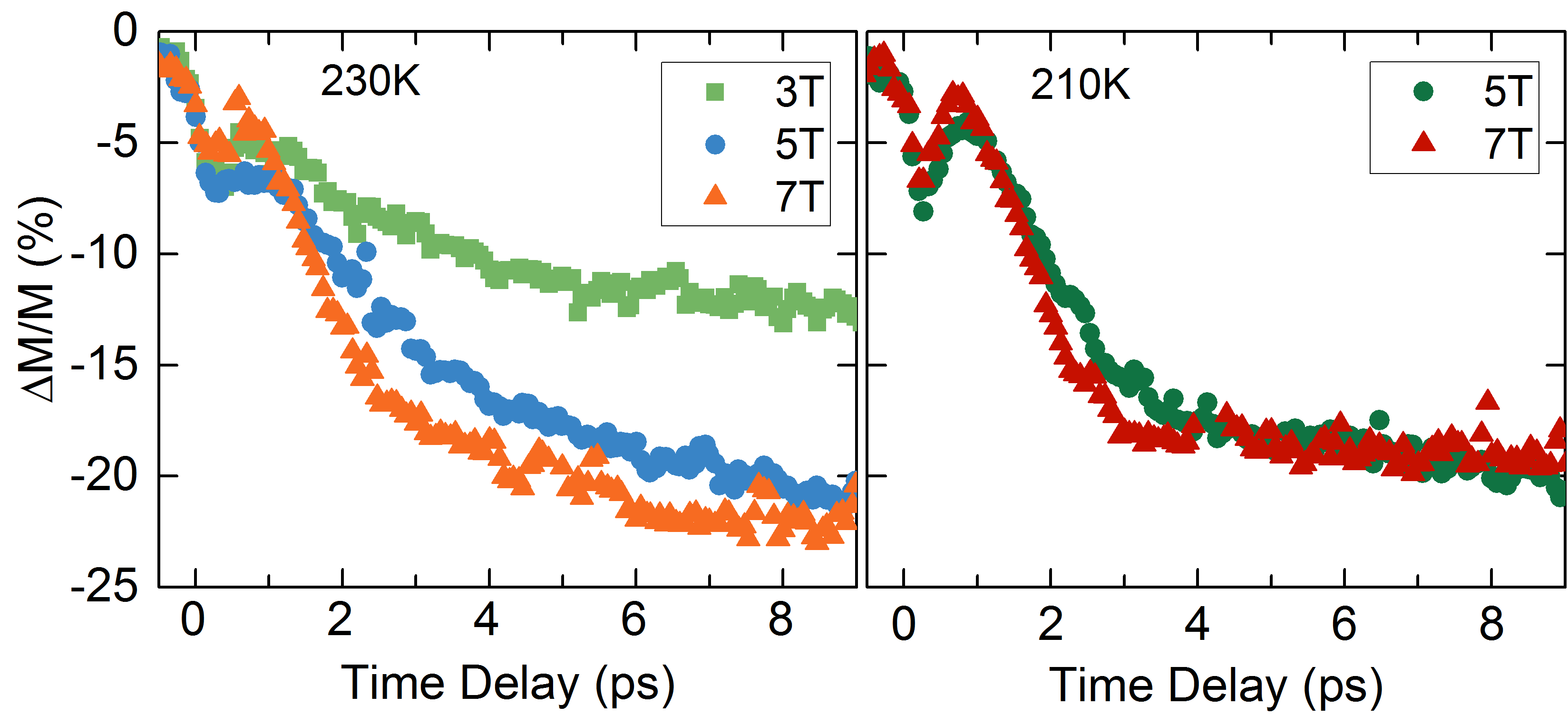}
	\caption{Demagnetization dynamics for three different fields at \SI{230}{\kelvin} (left) and \SI{210}{\kelvin} (right). Interestingly, the slow process depends on the applied field already at delay time $> \SI{1}{\pico\second}$, while the fast one (sub \SI{1}{\pico\second}) does not.}
	\label{fig:field-dep_mod}
\end{figure}

The dynamics during the first \SI{20}{\pico\second} show that the demagnetization is non-monotonic. In \figurename~\ref{fig:field-dep_mod} the first, field-independent, ultrafast demagnetization step happens immediately after the laser pulse, followed by a plateau or a small re-magnetization from \SIrange{1}{1.5}{\pico\second}. After this transient state, the sample continues to demagnetize further, but at a slower rate dependent on the applied field.

This behaviour clearly resemble the demagnetizing dynamics of Gd(FeCo)$_3$, where the two sublattices demagnetize at different speeds due to both their different magnetic moments and strongly different intra-sublattice exchange constants. In the case of \mrg\ the atomic moments of the Mn sublattices are almost equal\cite{Betto2015} and, in addition, the exchange constants (both intra- and inter-sublattice)\cite{Fowley2018} are considerably stronger\cite{Ostler2012,Mentink2012,Davies2019}.
Thus, the demagnetization rates for the two sublattices are expected to be similar, which is clearly contradicted by the observation of a non-monotonic demagnetization process.

In order to understand this, we note that for a strongly coupled ferrimagnetic system, the effect of a short laser pulse is different from that in a simple ferromagnet. Indeed, due to the strong antiferromagnetic inter-sublattice exchange, the total spin angular momentum can be conserved by passing it from one sublattice to the other. However, the temperature dependencies of the two are strongly different, thus an equal change of momentum corresponds to a larger variation in effective temperature for the $4c$ sublattice than the $4a$. We think this is the main reason for the emergence of a strongly non-equilibrium magnetic state in \mrg. 

In \figurename~\ref{fig:field-dep_mod} we also show the dependence of the demagnetization process on the magnitude of the applied magnetic field.
An increase of the external field leads to a faster dynamics only for the second, slower part of the process, while the first step of demagnetization remains unaffected.
During this second step of the demagnetization, the temperature of the electronic system is sufficiently reduced and does not dominate the overall dynamics of the magnetic system. That is why the influence of field becomes noticeable.
Between \SIrange{2}{4}{\pico\second} in right panel of \figurename~\ref{fig:field-dep_mod}, we tentatively note two periods of a high-frequency oscillation, $f \sim \SIrange[range-phrase = \text{~to~}]{0.6}{1}{\tera\hertz}$ which is in agreement with the frequency expected for the antiferromagnetic mode of \mrg.

In order to model the different behaviour of the two Mn-sublattices, we performed calculations using the 4TM described above: four coupled differential equations that describe the effect of a laser pulse on the different heat baths.
The interaction of the pump pulse with the sample is described as a sudden increase of the electronic temperature, $T_{e}$. Then, the system thermalizes by redistributing the heat to different heat-baths, with different time constants for each subsystem. The lattice is considered as a phonon bath ($T_{l}$), while the two magnetic sublattices $4a$ and $4c$ are represented by two different temperatures --- $T_{4a}$ and $T_{4c}$, respectively.

Initially, we used the set of $G$-parameters of GdCo/GdCoFe for the coupling constants.\cite{Beaurepaire1996,Seixas2010} For the specific heat capacities of lattice and electron systems, $C_{l}$ and $C_{e}$, we used values for Mn\textsubscript{2}Ga single crystals,\cite{Winterlik2008} shown in \tablename~\ref{table:4TM}. The solution of the system of equations thus gives us the time evolution of the temperatures of the different subsystems, i.e.\ electrons, lattice and the two spin sublattices.

\begin{table}
  \begin{ruledtabular}
	\begin{tabular}{l c c c}
		Constants & \mrx & GdCoFe & Unit \\%
		\hline 
		\noalign{\vskip 1mm}
		$C_{e}$ & \tablenum{484} & \tablenum{714} & \si{\joule\per\metre\cubed\per\kelvin\squared} \\ 
		$C_{l}$ & \tablenum{2.27e6} & \tablenum{3e6} & \si{\joule\per\metre\cubed\per\kelvin} \\ 
		$C_{4a/Gd}$ & \tablenum{2e5} & \tablenum{2.5e4} & \si{\joule\per\metre\cubed\per\kelvin} \\ 
		$C_{4c/CoFe}$ & \tablenum{0.3e4} & \tablenum{6e4} & \si{\joule\per\metre\cubed\per\kelvin} \\ 
		$G_{el}$ & \tablenum{8e17} & \tablenum{8e17} & \si{\joule\per\metre\cubed\per\kelvin} \\ 
		$G_{es}^{4c/CoFe}$ & \tablenum{2.8e15} & \tablenum{6e14} & \si{\watt\per\metre\cubed\per\kelvin} \\ 
		$G_{es}^{4a/Gd}$ & \tablenum{3e15} & \tablenum{1.4e16} & \si{\watt\per\metre\cubed\per\kelvin} \\ 
		$G_{ls}^{4c/CoFe}$ & \tablenum{3e15} & \tablenum{3e14} & \si{\watt\per\metre\cubed\per\kelvin} \\ 
		$G_{ls}^{4a/Gd}$ & \tablenum{9e16} & \tablenum{3e15} & \si{\watt\per\metre\cubed\per\kelvin} \\ 
		$G_{ss}$ & \tablenum{2.8e16} & \tablenum{1.6e15} & \si{\watt\per\metre\cubed\per\kelvin} \\ 
	\end{tabular}
      \end{ruledtabular}
	\caption{Parameters used for the 4TM. The second column shows values for GdCoFe\cite{Mekonnen2013}.}
	\label{table:4TM}
\end{table}

Fitting a single set of experimental observations with ten unknown parameters is not a recipe for reliable results. We therefore adjust each parameter manually to obtain a good agreement, while restricting them to \emph{realistic} values.
A difference in the electronic heat capacity of the two sublattices, $C_{4a}$ and $C_{4c}$, is expected due to their different electronic density of states. In addition, the coupling constants can be qualitatively related to the strength of the magnetic exchange. We expect a strong spin-spin coupling ($G_{ss}$) and a stronger lattice-spin coupling for the $4a$ sublattice, $G_{ls}^{4a}$, compared to the $4c$ one.

\begin{figure}
	\centering
	\includegraphics[width=\columnwidth]{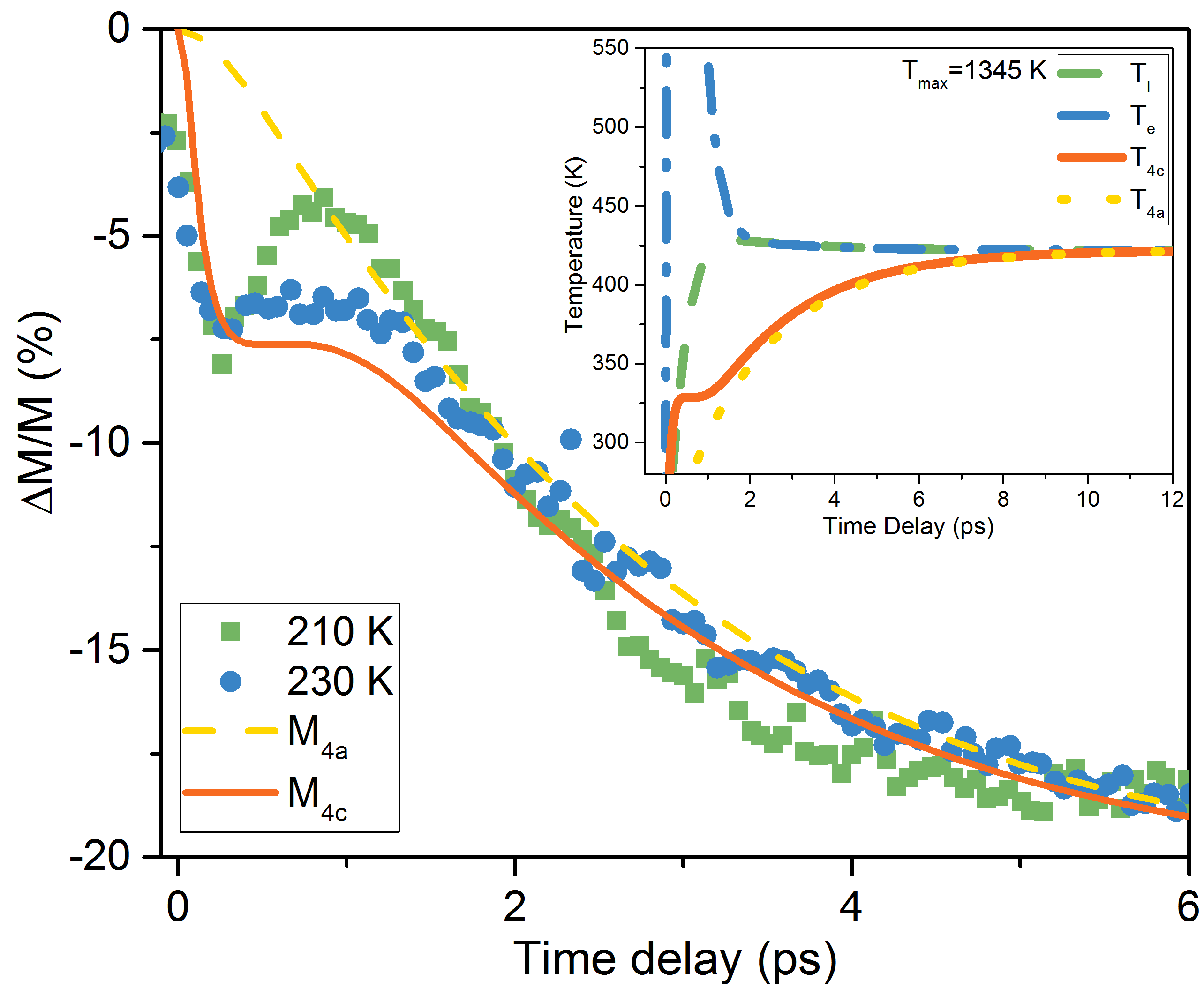}
	\caption{Laser-induced demagnetization in an applied field of \SI{5}{\tesla}. Thick lines are curves obtained from the 4TM. Inset: Time dependence of electron, lattice, $4c$ and $4a$ heat reservoirs. The electronic temperature increases above $T_{C}$ following the laser excitation, and thermalize within \SI{1}{\pico\second}. It reaches equilibrium with the lattice within \SI{2}{\pico\second}. The temperature of the $4c$ and $4a$ sublattices strongly differ for the first \SI{2}{\pico\second}, after which they are in equilibrium.}
	\label{fig:two-step}
\end{figure}

Using the adjusted parameters, the electronic temperature, $T_{e}$ (\figurename~\ref{fig:two-step} (inset)), reaches \SI{1345}{\kelvin} in roughly \SI{100}{\femto\second}, while the lattice and sublattices $4c$/$4a$ ($T_{l}$, $T_{4c}$ and $T_{4a}$ respectively) remain close to \SI{350}{\kelvin}. The heat deposited in the electronic system is then redistributed between the other subsystems. In particular, within \SI{2}{\pico\second} the electronic and lattice subsystems are in thermal equilibrium. On the other hand, for the two spin subsystems, full thermal equilibrium is only reached  after $\sim \SI{10}{\pico\second}$. This behaviour is quite similar to that of GdCoFe, but with one major difference. We note that the temperature of the two spin subsystems in \mrg\ follow a similar relaxation path already at $\sim \SI{2}{\pico\second}$, while for GdFeCo the relaxation times are quite different. As explained above, this suggests that the interplay of a strong exchange coupling between sublattices (inter-exchange) and the electronic structure of \mrg\ leads to dynamics where the total spin angular momentum of the two sublattices is practically conserved after a very short time of $\sim \SIrange[range-phrase = \text{ to }]{1}{2}{\pico\second}$.

To compare with experimental data, we converted the temperature-time dependencies following a $T^{3/2}$ Bloch law\cite{Beaurepaire1996,Mekonnen2013} as the strong exchange keeps some amount of magnetic order even in the non-equilibrium state (see above). \figurename~\ref{fig:two-step} shows reasonable agreement between experimental data and the 4TM\@. In addition to the experimental data representing the $4c$ sublattice magnetization, we also show the $4a$ magnetization, inferred from the model, to highlight its strong influence on the demagnetization process. What we observe is an ultrafast demagnetization of one of the sublattices (assumed to be $4c$), followed by a secular equilibrium (when the temperature of the measurement is \SI{230}{\kelvin}) or by a fast re-magnetization (at \SI{210}{\kelvin}), and only after $\sim \SI{1.5}{\pico\second}$ does the second sublattice start to demagnetize, reaching its minimum after $\sim \SI{20}{\pico\second}$.

Regarding the refined values of the 4TM parameters, we highlight two points. First, the coupling constant of the two magnetic sublattices is considerably stronger in \mrg\ than for GdCoFe, as expected given the higher exchange coupling. Second, a strong difference is found in the heat capacity of the two sublattices. These values are in line with what could be expected from \mrg\ with its two different manganese spin systems.


In conslusion, we have shown that a femtosecond pump pulse can demagnetize \mrg\ in approximately ten picoseconds via a two-step process. This result is similar to what was already observed for amorphous GdCoFe alloys\cite{Mekonnen2013}.
Surprisingly, here we observe a faster evolution of the demagnetization dynamics. Indeed, one of the sublattices (assumed to be $4c$, based on earlier experiments and density functional theory\cite{Zic2016}) demagnetizes in few hundred of \si{\femto\second}, and at $\sim \SI{1.5}{\pico\second}$ a second demagnetization process starts, that is assigned to the second sublattice ($4a$).
We underline that the process observed here, and the apparent faster demagnetization of one sublattice, arises from the exchange-driven dynamics. This is supported by the similar demagnetization rate of Mn in the two sublattices and by the strong exchange in \mrg\@.

We have modelled the experimental data, using the phenomenological 4TM model, thereby establishing, at least approximately, the intrinsic properties that govern not only demagnetization but also all-optical switching. We stress that, even though we only observe a partial demagnetization, these results highlight a pathway towards all-optical-switching in ferrimagnetic Heusler alloys. A faster demagnetization rate is essentially connected to faster heat-transfer and smaller heat capacity, that can lead to deterministic all-optical switching of \mrg\ with switching times as short as $\sim \SI{1}{\pico\second}$ when the two spin reservoirs achieve equilibrium.

\begin{acknowledgments}
	This project has received funding from the NWO, the \emph{European Union's} Horizon 2020 research and innovation programme under grant agreement No 737038 `TRANSPIRE', as well as from Science Foundation Ireland through contracts 12/RC/2278 AMBER and 16/IA/4534 ZEMS
\end{acknowledgments}

\bibliography{refarticleDEMAG}

\end{document}